\pdfoutput=1
\documentclass{aa}  
\usepackage{amsmath}
\usepackage{euscript}
\usepackage{balance}
\usepackage{graphicx}
\usepackage{txfonts}
\usepackage{color}
\usepackage[utf8]{inputenc}

\begin{document} 

   \title{Starspot signature on the light curve}
   \subtitle{Learning about the latitudinal distribution of spots}

   \author{A. R. G. Santos \inst{1,2,3}
          \and M. S. Cunha \inst{1,2}
          \and P. P. Avelino \inst{1,2}
          \and R. A. Garc\'{i}a \inst{4}
          \and S. Mathur \inst{5}}

   \institute{Instituto de Astrof\'{i}sica e Ci\^{e}ncias do Espa\c{c}o, Universidade do Porto, CAUP, Rua das
Estrelas, PT4150-762 Porto, Portugal\\
              \email{asantos@astro.up.pt}
              \and Departamento de F\'{i}sica e Astronomia, Faculdade de Ci\^{e}ncias, Universidade do Porto,
Rua do Campo Alegre 687, PT4169-007 Porto, Portugal
              \and School of Physics and Astronomy, University of Birmingham, Edgbaston, Birmingham B15 2TT, UK
              \and Laboratoire AIM, CEA/DRF-CNRS-Universit\'{e} Paris 7 Diderot; IRFU/SAp, Centre de Saclay, 91191 Gif-sur-Yvette Cedex, France
              \and Center for Extrasolar Planetary Systems, Space Science Institute, 4750 Walnut street Suite\#205, Boulder CO 80301, USA
}

  \date{\today}

  \abstract
{Quasi-periodic modulations of the stellar light curve may result from dark spots crossing the visible stellar disc. Due to differential rotation, spots at different latitudes generally have different rotation periods. Hence, by studying spot-induced modulations, one can learn about stellar surface (differential) rotation and magnetic activity. Recently, Reinhold \& Arlt (2015) proposed a method based on the Lomb-Scargle periodogram of light curves to identify the sign of the differential rotation at the stellar surface.}
{Our goal is to understand how the modulation of the stellar light curve due to the presence of spots and the corresponding periodogram are affected by both the stellar and spot properties.}
{We generate synthetic light curves of stars with different properties (inclination angle, limb darkening, and rotation rate) and spot configurations (number of spots, latitude, intensity contrast, and size). By analysing their Lomb-Scargle periodograms, we compute the ratio between the heights of the second and first harmonics of the rotation period (peak-height-ratio).}
{We find that the peak-height ratios are essentially a function of a single parameter, the fraction of time the spot is visible, which is related to the sinusoidality of the spot modulation. We identify the conditions under which the periodogram analysis can actually provide an estimate of the spot latitudes and/or the stellar inclination angle. We also identify possible sources of error in the identification of the sign of the differential rotation.}
{}


  \maketitle
%

\section{Introduction}

Stellar rotation, in particular differential rotation, is a key ingredient of the dynamo mechanism, which is responsible for the generation of the magnetic field in the Sun and solar-like stars. As a manifestation of the magnetic activity, dark spots emerge at the stellar surface. Spots are regions of strong magnetic field that suppresses the convection, resulting on a less efficient heat transport. Therefore, spots are cooler and, consequently, darker than the surroundings, having an impact on the stellar brightness. As the spots cross the stellar visible disc, they modulate the light curve. Such modulation provides information about the stellar rotation and magnetic activity \citep[e.g.][]{Mosser2009,Mathur2010,Garcia2010a,Ballot2011,Garcia2014,Mathur2014}.

High-precision photometric time series obtained with space telescopes allowed the detection of rotational periods for a large number of stars \citep[e.g.][]{Reinhold2013a,Nielsen2013,McQuillan2013,McQuillan2013a,McQuillan2014,Garcia2014}, through methods based on the Lomb-Scargle periodogram, the autocorrelation function and/or the wavelet transform. Moreover, the high quality of these time series provides a good opportunity to measure differential rotation, since the spot-induced modulations of the light curves enclose specific signatures of spots at different latitudes. The amplitude of the differential rotation can be recovered through spot modelling \citep[e.g.][]{Mosser2009,Huber2010,Lanza2011,Lanza2014}. By fitting a given model to the observed light curve, a number of stellar and spot parameters may be constrained, including the spots' rotation rates. 
Differential rotation has also been measured through the periodogram analysis \citep[e.g.][]{Reinhold2013,Reinhold2013a,Reinhold2015a,Nagel2016,Distefano2016a}. In the periodogram, broad or multiple peaks associated to the stellar rotation are usually interpreted as evidence of the differential rotation. The analysis of individual subseries of the full light curve, whose modulation might be dominated by a given spot at a given latitude, allows the identification of temporal variations in the recovered rotation period, which can also be an indication of differential rotation.

Recently, \citet{Reinhold2015} proposed a new method, based on the periodogram analysis, to detect the sign of differential rotation, that is, whether the equatorial regions rotate faster ($+$, solar differential rotation) or slower ($-$, antisolar differential rotation) than the poles. When they apply the method to a particular set of synthetic light curves with solar differential rotation, a low false-positive rate ($11.3\%$--$20\%$) was recovered. The method was also applied to a sample of 50 stars observed by {\it Kepler}. Solar differential rotation was reported for 21--34 stars, while 5--10 stars were found to be consistent with anti-solar differential rotation \citep[for details, see][]{Reinhold2015}.

In this work, we investigate the spot's signature on the light curve and, consequently, on the periodogram. In particular, we are interested in understanding the conditions that lead to the successful or unsuccessful detection of the sign of differential rotation. 

\section{Method to determine the sign of the surface differential rotation}\label{sec:method}

The Lomb-Scargle periodogram (LSP) may be used to determine the stellar surface rotation. Secondary peaks close to the main rotation period are interpreted as evidence for differential rotation, being associated with spots/active regions at different latitudes, thus rotating at different rates. 

\citet{Reinhold2015} proposed a new and simple method to determine the sign of differential rotation, which consists in comparing the ratios between the height of the second and first harmonics (hereafter peak-height ratios) associated to different rotation periods. For a given rotation period $P_j$ (first harmonic), the peak-height ratios are computed as 
\begin{equation}
r_j=\dfrac{h(P_j')}{h(P_j)},
\end{equation}
where $P_j'$ is the second harmonic of the rotation period, and $h(P_j)$ and $h(P_j')$ are the heights of the first and second harmonics, respectively (hereafter, $h_j$ and $h_j'$).

The authors argued that spots at lower latitudes lead to  less sine-shaped light curves than spots at higher latitudes, resulting in extra power on the second harmonic and, thus, larger peak-height ratios. While they based their argument on results for synthetic light curves with specific configurations, the authors do not provide any further explanation. In Sect. \ref{sec:1-spot}. we will address in detail the latitudinal dependence of the peak-height ratios.

Following their argument that spots at lower latitudes lead to larger peak-height ratios than spots at higher latitudes, when comparing the peak-height ratios of two periods associated with the surface rotation, $P_j$ and $P_{j+1}$, the method allows for the determination of a relative latitude ("high" or "low") and, thus, the sign of differential rotation. \citet{Reinhold2015} define the observed relative differential rotation as
\begin{equation}
\alpha_{\rm obs}=\dfrac{P_{\rm high}-P_{\rm low}}{P_{\rm high}}.\label{eq:alpha}
\end{equation}
$\alpha_{\rm obs}>0$ corresponds to solar differential rotation (the equator rotates faster than the poles) and $\alpha_{\rm obs}<0$ corresponds to antisolar differential rotation (the poles rotate faster than the equator).

\section{Synthetic light curves}\label{sec:toollc}

In order to study the modulation of the stellar light curves due to the presence of spots, we developed a tool to simulate the light curves of spotted stars based on the models of \citet{Lanza1993} and \citet{Eker1994}.

Each spot is assumed to be circular and is decomposed in a number of area elements. The total decrease in flux due to spots corresponds to the sum of the contributions from each element $k$
\begin{equation}
\dfrac{\Delta F}{F}=\sum_k\left(\dfrac{\Delta F}{F}\right)_k.
\end{equation}
The decrease in flux associated to an element $k$ is given by
\begin{equation}
\left(\dfrac{\Delta F}{F}\right)_k=(1-C_{\rm S})\dfrac{S_{\!k}}{\pi R^2_*}\mu_k\dfrac{I(\mu_k)}{I(1)},
\end{equation}
where $S_{\!k}$ is the element area, $R_*$ is the stellar radius, $I(\mu_k)/I(1)$ is the relative photospheric intensity given by the limb-darkening law, and $\mu_k=\cos \psi_k$ where $\psi_k$ is the angle between the line of sight and the normal to the surface element given by
\begin{equation}
\mu_k=\cos\psi_k=\cos i\cos\theta_k+\sin i\sin\theta_k\cos\phi_k.
\end{equation}
Here, $i$ is the stellar inclination angle, i.e. the angle between the stellar rotation axis and the line of sight, and $\theta_k$ and $\phi_k$ are the colatitude and longitude of the element. The element $k$ is visible whenever $0\leq\mu_k\leq1$. $C_{\rm S}$ is the spot-to-photosphere intensity ratio, i.e. $C_{\rm S}\!=\!I_{\rm S}/I_{\rm P}$, which, for simplicity, we shall assume to be a constant.

\section{Results}\label{sec:results_syn}

\subsection{1-spot simulations}\label{sec:1-spot}

The modulation on the light curve induced by spots crossing  the visible disc of the star depends on a number of stellar and spot parameters, e.g. the stellar inclination angle, rotation rate, limb-darkening law, spot size, latitude, and contrast.

We start by investigating to what extent the peak-height ratios, $r=h'/h$, are a measure of the sinusoidality of the spot modulation on the light curve. To do so, we shall start by considering the simplest case of 1-spot simulations. We obtain the synthetic light curves for stars with different inclination angles, $i$, and a single spot at different latitudes, $L$. For this set of simulations we assume a circular spot of constant radius $R_{\rm S}\sim5.7^\circ$ ($A=5000\,\mu\rm Hem$; about the area covered by sunspots at solar maximum), infinite lifetime and an intensity contrast of $C_{\rm S}=0.67$ \citep[e.g.][]{Sofia1982,Lanza2003,Walkowicz2013}. Also, we consider a quadratic limb-darkening law
\begin{equation}
\dfrac{I(\mu)}{I(1)}=1-a(1-\mu)+b(1-\mu)^2,\label{eq:qld}
\end{equation}
where we have assumed $a=0.5287$ and $b=0.2175$, which are adequate for solar-like stars observed by {\it Kepler} \citep[][]{Claret2000,Reinhold2013}. The differential rotation is assumed to be solar and is given by
\begin{equation}
\Omega(L)=\Omega_{\rm eq}(1-\alpha\sin^2L-\beta\sin^4L),\label{eq:snodgrass}
\end{equation}
where $\Omega_{\rm eq}$ is the angular velocity at the equator, and $\alpha$ and $\beta$ are the parameters that determine the latitudinal dependency of the rotation rate. For this set of synthetic data, we have considered $\Omega_{\rm eq}=0.2567\,\rm rad\, d^{-1}$, $\alpha=0.1584$
, and $\beta=0.1210$ \citep[][]{Snodgrass1983,Snodgrass1990}. The initial longitude of the spot for each simulation in this section is determined randomly.

\begin{figure}[h]
\includegraphics[width=\hsize]{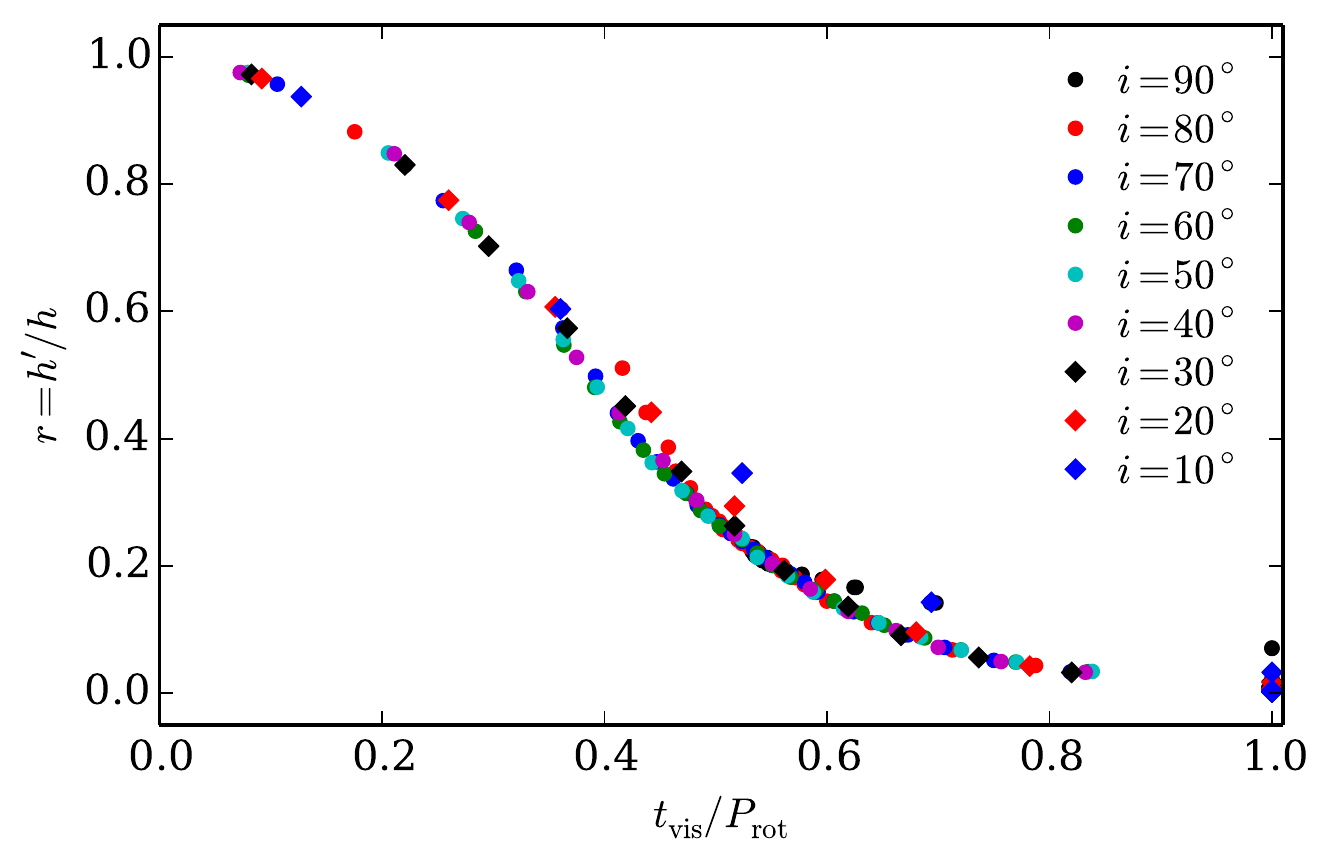}
\caption{Peak-height ratios associated to spots at different latitudes (ranging from $0^\circ$ to $\pm85^\circ$) as a function of the spot visibility time. Different colors indicate different stellar inclination angles, $i$.}\label{fig:ratioLtv}
\end{figure}

We compute the Lomb-Scargle periodogram for each synthetic light curve and the corresponding ratios between the second and first harmonics. 
We find that the peak-height ratios are essentially a function of a single parameter: the visibility time of the spot. Figure \ref{fig:ratioLtv} shows how the peak-height ratios change as a function of the ratio between the visibility time and the rotation period. Spots that are visible for most of the rotation period lead to more sine-shaped signals than spots that are visible for a smaller fraction of time. The spot is considered visible whenever there is a decrease in flux. Using this definition for the visibility time of the spot ($t_{\rm vis}$), we might be overestimating the true visibility time, specially for low inclination angles. Nevertheless, we can clearly conclude that the longer the spots are visible the smaller the peak-height ratios are.

\begin{figure}
\includegraphics[width=\hsize]{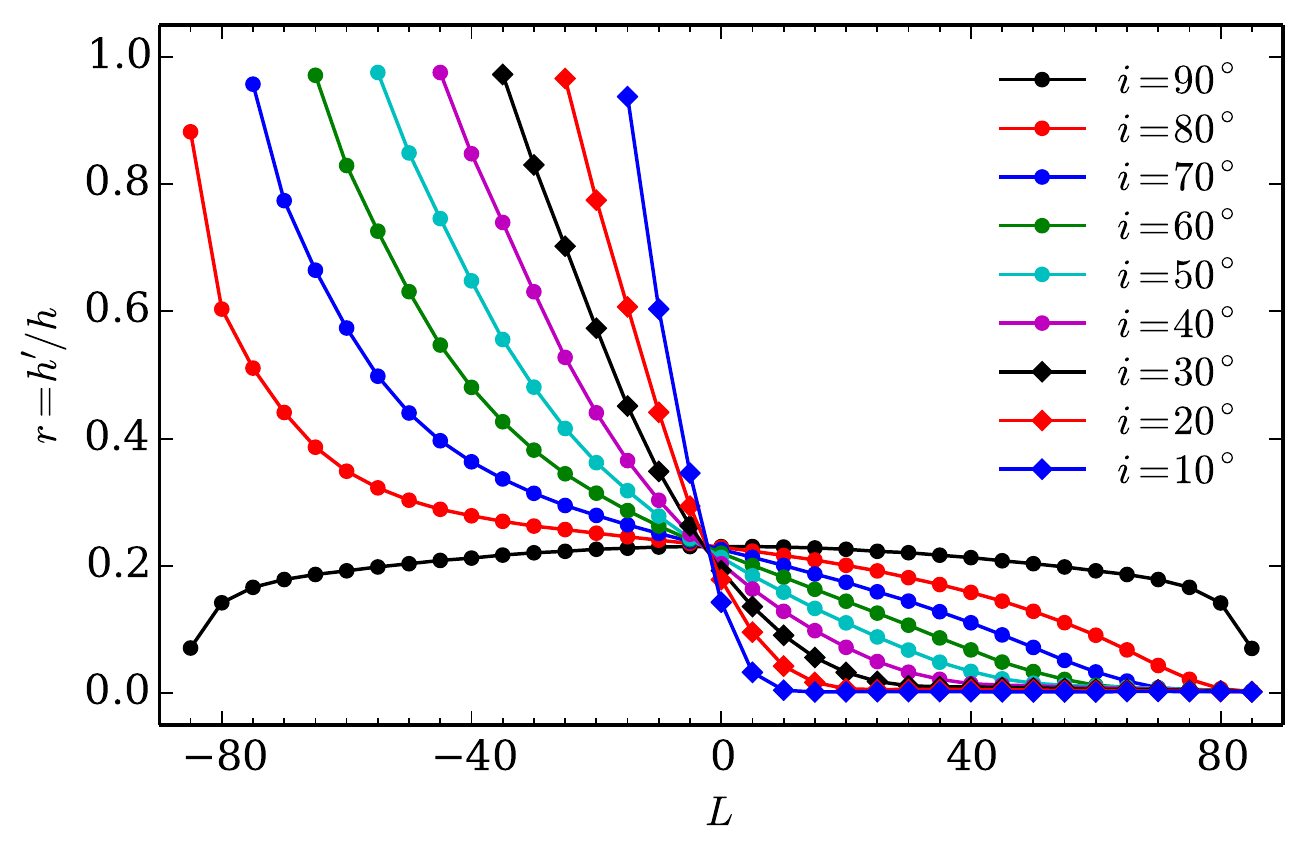}
\caption{Peak-height ratios as a function of the spot latitude, $L$, for different inclination angles, $i$.}\label{fig:ratioL}
\end{figure}

Figure \ref{fig:ratioL} shows the peak-height ratios as a function of the spot latitude, which together with the stellar inclination angle are the most determinant parameters for the spot visibility. As different combinations of $i$ and $L$ result on the same spot visibility time, there is a degeneracy between latitude and inclination. Nevertheless, the peak-height ratios provide constrains on the possible solutions $(i,L)$ that can lead to the spot signature on the observed light curve. If the stellar inclination is known one can estimate the spot latitude.

The method proposed by \citet{Reinhold2015} for the determination of the sign of the surface differential rotation relies on the correct identification of a relative latitude ("low"/"high") for at least two rotation periods (first harmonics) in the LSP. Implicit to their method is the assumption that spots at lower absolute latitudes are associated with higher peak-height ratios, $r=h'/h$, than spots at higher absolute latitudes. Except for inclination angles very close to $i=90^\circ$, this is only true for spots on the same hemisphere as the observer (that we will call northern hemisphere). Spots emerging at higher latitudes on the southern hemisphere are visible for a smaller fraction of time, thus, inducing a less sinusoidal signature and leading to higher peak-height ratios than spots at lower latitudes on the northern or southern hemispheres. Hence, for values of the inclination angle not too close to $i=90^\circ$, the method will suggest the wrong sign for the differential rotation when comparing the peak-height ratios of periods associated to spots on the southern hemisphere. The wrong sign will also be recovered when one of the spots is at $L_1$ on the northern hemisphere, the second is at $L_2$ on the southern and $|L_2|\!>\!|L_1|$.

For an inclination angle of $i=90^\circ$, the behaviour of the peak-height ratios is hemispheric symmetric and nearly independent on the latitude of the spot (except for $|L|$ very close to $90^\circ$). Therefore, for this inclination, the association of the detected rotation periods to different latitudes will be difficult. Also, for small inclination angles, the ratios become saturated at high latitudes on the northern hemisphere, as spots at that location are always visible.

Although our results show that the peak-height ratios are essentially a function of the visibility time of the spot, which is determined mainly by the stellar inclination angle and the spot latitude, the modulation in the light curves induced by spots also depends on other parameters. In what follows, we investigate the impact on the peak-height ratios of other spot and stellar properties, such as the spot area and relative intensity, rotation rate and limb-darkening law.

\begin{figure*}[!]
\includegraphics[width=\hsize]{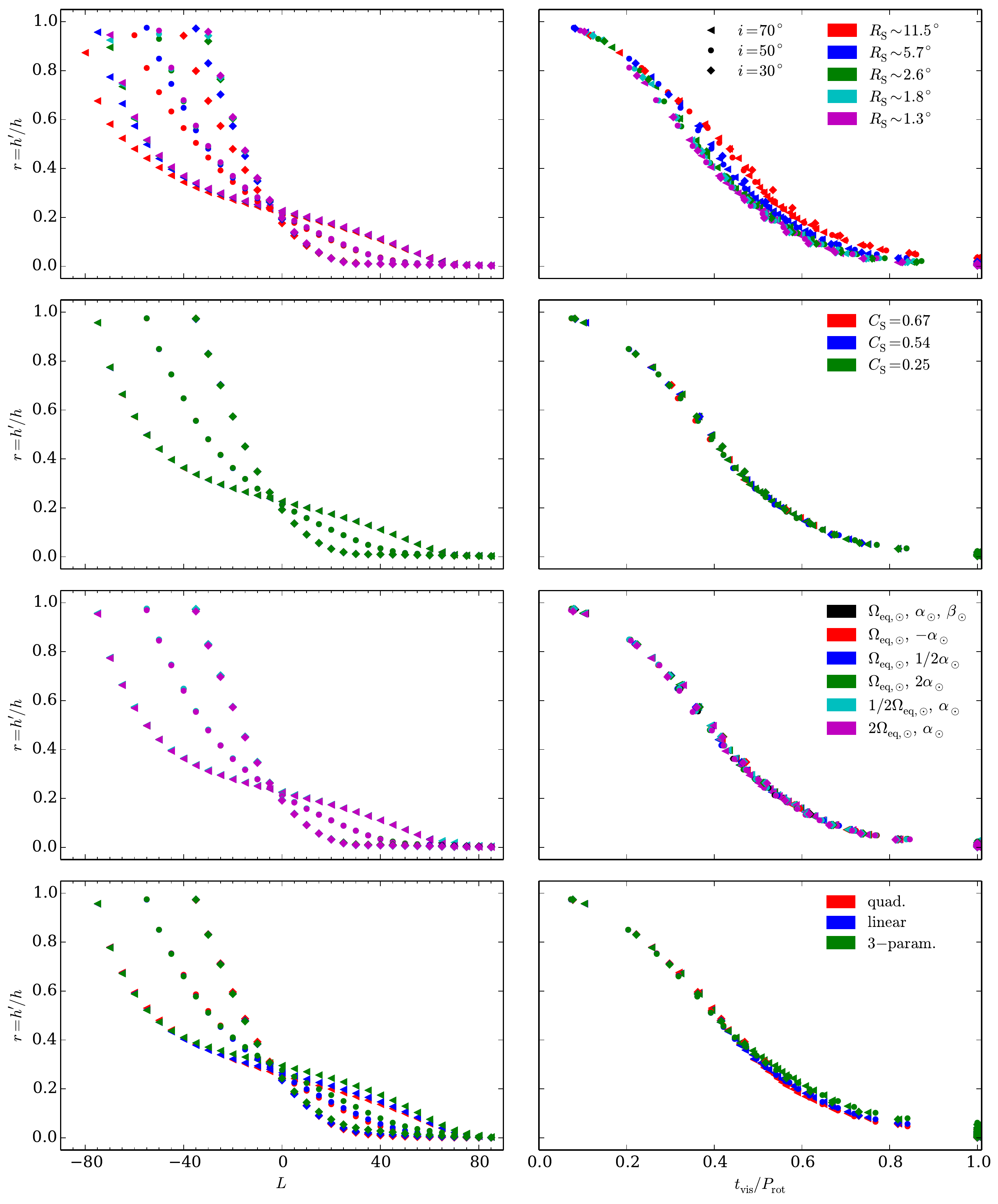}
\caption{Peak-height ratios as a function of the spot latitude (left) and visibility time (right) for the inclination angles $i=30^\circ$, $50^\circ$, and $70^\circ$  and for different spot areas (top panel), different spot-to-photosphere intensity ratios (second row), various differential rotation parameters (third row), and different limb-darkening laws (bottom panel). For these simulations, the default values of the spot radius and intensity contrast are $R_{\rm S}\sim5.7^\circ$ and $C_{\rm S}=0.67$l. We use the solar rotation ($\Omega_{\rm eq, \odot}$, $\alpha_{\odot}$, and $\beta_{\odot}$) as the default rotation profile. Finally, the default limb-darkening law is the quadratic one with parameters $a=0.5287$ and $b=0.2175$. [$R_{\rm S}\!\sim\!11.5^\circ\Leftrightarrow A_{\rm S}=\!20000 \,\mu{\rm Hem}$, $R_{\rm S}\!\sim\!5.7^\circ\Leftrightarrow A_{\rm S}=\!5000 \,\mu{\rm Hem}$, $R_{\rm S}\!\sim\!2.6^\circ\Leftrightarrow A_{\rm S}=\!1000 \,\mu{\rm Hem}$, $R_{\rm S}\!\sim\!1.8^\circ\Leftrightarrow A_{\rm S}=\!500 \,\mu{\rm Hem}$, $R_{\rm S}\!\sim\!1.3^\circ\Leftrightarrow A_{\rm S}=\!250\,\mu{\rm Hem}$]}\label{fig:ratiopar}
\end{figure*}

The top panel of Fig. \ref{fig:ratiopar} shows the peak-height ratios as a function of latitude (left panel) and visibility time (right panel) for different inclination angles and spot sizes. The impact of the spot size on the recovered ratios is more significant for spots at higher latitudes on the southern hemisphere and lower inclinations. For a given latitude and inclination, a larger spot will be visible for longer than a smaller spot, thus larger spots lead to smaller ratios than smaller spots. 

As the spot-to-photosphere intensity contrast does not affect the visibility time of the spot, it also does not have a significant impact on the peak-height ratios. This is shown in the second row of Fig. \ref{fig:ratiopar}. As mentioned before, for these synthetic light curves the initial phase of the spot is determined randomly, which together with the fact that the light curve is discrete, introduces a small effect on the estimated spot visibility time. The small differences seen in the right panel of the second row in Fig. \ref{fig:ratiopar} show that the phase of a given spot alone has little impact on the visibility time of the spot.

In order to investigate the impact of the rotation rate on the peak-height ratios, we have considered different rotation profiles in the synthetic data, including solar ($\alpha>0$) and anti-solar ($\alpha<0$) differential rotation. In this set of simulations, we consider the simplified version of Eq. (\ref{eq:snodgrass}) that is commonly used,
\begin{equation}
\Omega(L)=\Omega_{\rm eq}(1-\alpha\sin^2L).\label{eq:rot2}
\end{equation}
The third row of Fig. \ref{fig:ratiopar} summarizes the results from this study, where $\Omega_{\rm eq,\,\odot}=0.2567\,\rm rad\, d^{-1}$ and $\alpha_\odot=0.1584$ denote the solar values considered above. Since the rotation rate does not change the fraction of time the spot is visible, it does not affect significantly the peak-height ratios. However, small discrepancies are still visible, which result first from the random initial spot phases, and second from the fact that while the characteristic time-scale of the light curves changes when considering different rotation rates, the length and cadence of the light curves are unchanged.

The bottom panel of Fig. \ref{fig:ratiopar} shows the results obtained from synthetic data considering different limb-darkening laws: the quadric limb-darkening law (Eq. (\ref{eq:qld})), the linear limb-darkening law
\begin{equation}
\dfrac{I(\mu)}{I(1)}=1-u(1-\mu),
\end{equation}
and the 3-parameter non-linear limb-darkening law
\begin{equation}
\dfrac{I(\mu)}{I(1)}=1-c_2\left(1-\mu\right)-c_3\left(1-\mu^{3/2}\right)-c_4\left(1-\mu^2\right),
\end{equation}
where $u$, $c_2$, $c_3$, and $c_4$ are the limb-darkening coefficients, which we take from the study by \citet{Sing2010} for {\it Kepler} data. Since the effective temperature ($T_{\rm eff}$), the surface gravity ($\log\,g$), and metallicity ($[\rm M/H]$) are, in principle, known parameters, Fig. \ref{fig:ratiopar} shows the results for $T_{\rm eff}=5750\,\rm K$, $\log\,g=4.50$ and $[\rm M/H]=0.00$. As the limb-darkening changes the shape of the spot modulation, it also affects the sinusoidality of the modulation seen through the peak-height ratios (bottom panel of Fig. \ref{fig:ratiopar}). Also, for different inclinations, spots with the same $t_{\rm vis}/P_{\rm rot}$ have different trajectories over the visible disc, corresponding to different limb-darkening and projected spot areas. In turn, the sinusoidality of the spot signature changes. This effect is small and can be seen through the differences between different inclinations (for example, second row of Fig. \ref{fig:ratiopar}).

\subsection{2-spot simulations}\label{sec:2-spot}

In this section, we analyse synthetic light curves obtained considering two spots on the stellar surface, in a broad range of latitudes, and we explore possible sources for contamination of the peak-height ratios. 

For the first set of synthetic light curves, the rotation rate is defined by Eq. (\ref{eq:rot2}) with parameters $\Omega_{\rm eq,\,\odot}$ and $\alpha_\odot$, and the spot radius is fixed at $5.7^\circ$. The length of the synthetic light curves is four years, consistent with the typical length of the {\it Kepler}'s light curves. The two spots have the same longitude at the beginning of the simulation. 

\begin{figure}[h]
\includegraphics[width=\hsize]{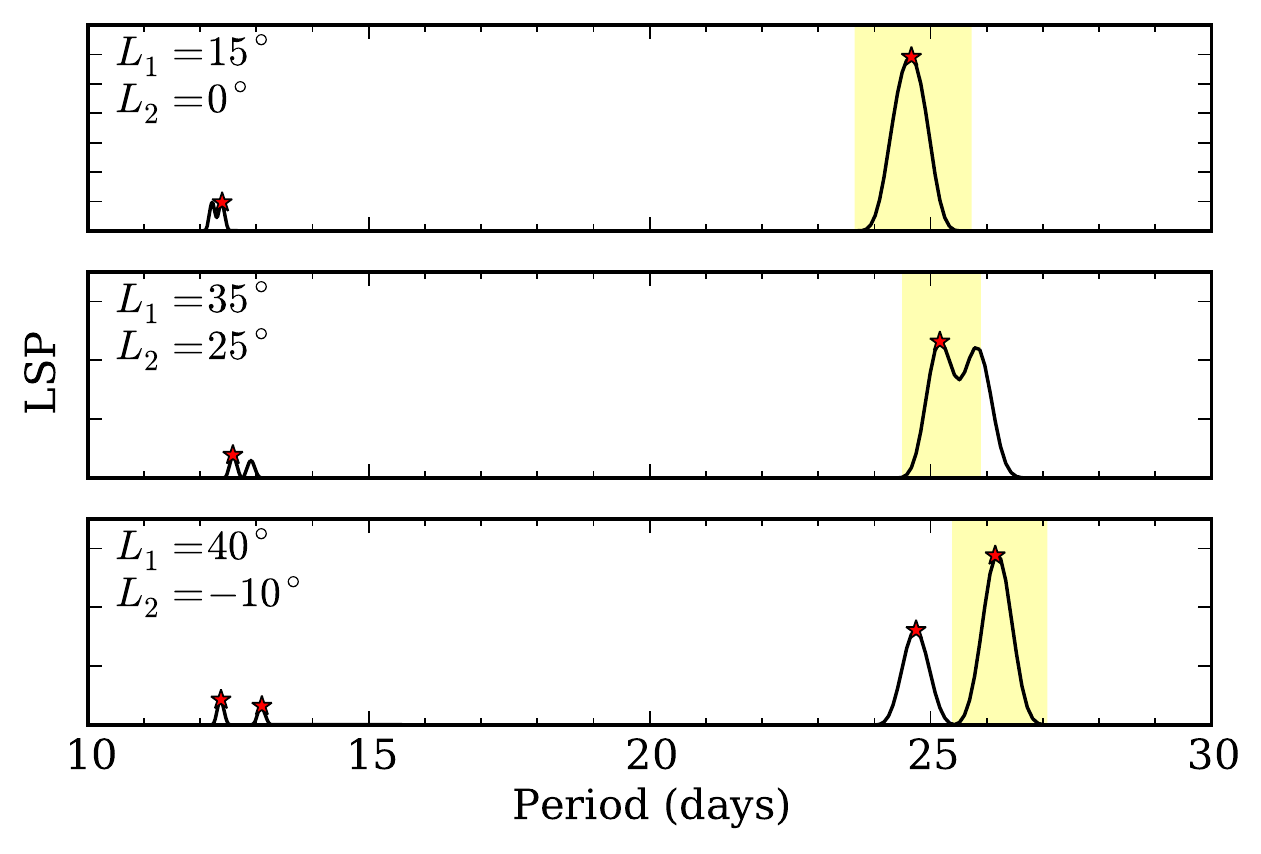}
\caption{Lomb-Scargle periogram for three synthetic light curves from simulations with two spots at different latitudes, $L_1$ and $L_2$. Top panel: Only one rotation period is detected. Middle panel: Two peaks associated to the surface rotation are seen but the second is not within the detectable period range. Bottom panel: The rotation periods associated to each spot latitude are successfully detected. The spot latitudes are indicated on the left top of each panel. The red symbols mark the first  and second harmonics if detected. The yellow regions mark the detectability limit we impose (see text for details).}\label{fig:lspseparation}
\end{figure}

The analysis performed in Sect. \ref{sec:1-spot} will only be valid in cases for which at least two rotation periods are clearly detected in the Lomb-Scargle periodogram. As peaks in the LSP can interfere with each other, we impose a detectability limit for the period separation. Because of the non-infinite light curve, we fit a sinc function in frequency (being symmetric in frequency, not in period) to the main peak, $P_1$, and define the detectability limit to be equal to 1.5 times the width of the sinc function at half maximum. Figure \ref{fig:lspseparation} shows the LSP for three different cases: i) a case where despite having two spots at different latitudes one is only able to recover one rotation period (top panel), ii) a case where two rotation periods might be recovered but they do not fulfil the chosen criteria on the minimum distance between two peaks (middle panel), and iii) a case where two rotation periods are clearly detected being separated by more than the imposed limit (bottom panel). Also, we discard peaks that may be significantly affected by the side lobes related to the first period.

\subsubsection{Spots' latitude effect}

\begin{figure*}[t!]
\includegraphics[width=\hsize]{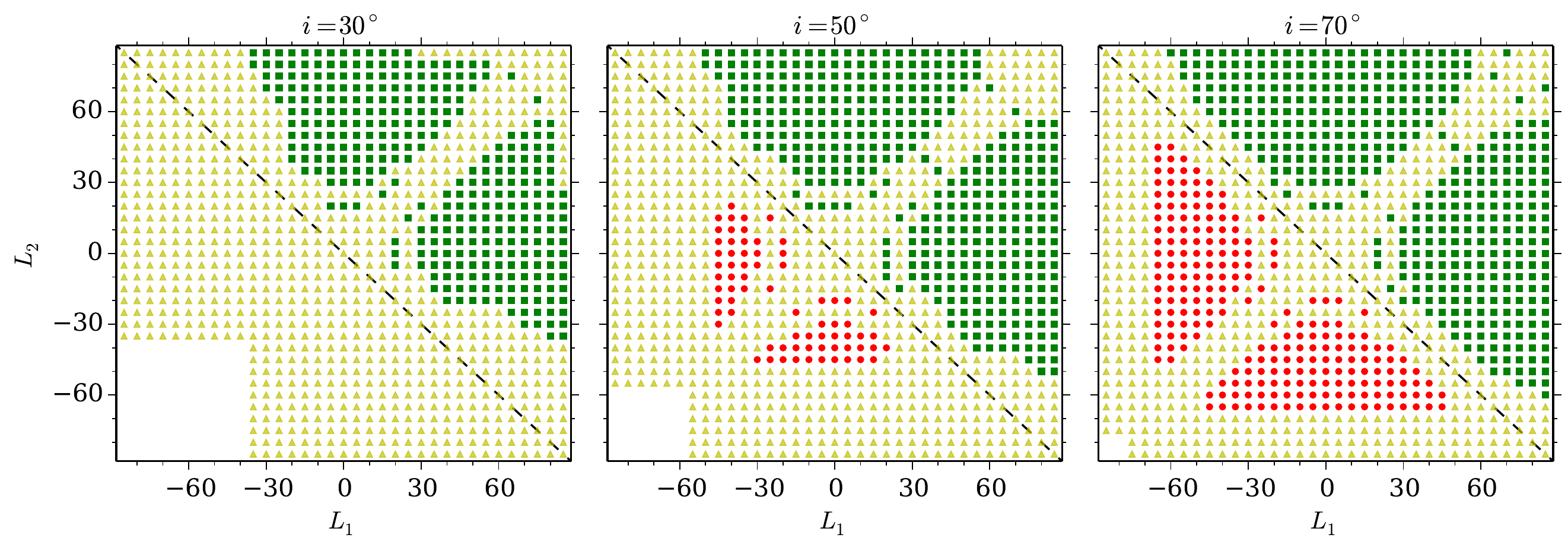}
\caption{Sign of the surface differential rotation, $\alpha_{\rm obs}$, for 2-spot simulations with stellar inclination angles of $i=30^\circ$ (left), $i=50^\circ$ (middle), and $i=70^\circ$ (right). The spot latitudes ($L_1$ and $L_2$) range from $-85^\circ$ and $85^\circ$ with steps of $5^\circ$. The yellow triangles represent the cases where only one rotation period is detected according to the criteria explained above. The red dots represent the cases where the wrong sing of differential rotation ($\alpha_{\rm obs}<0$) is found, while the green squares mark the cases where the correct sing ($\alpha_{\rm obs}>0$) is recovered. The dashed line divides the regions where the correct (above) or wrong (bellow) sign of $\alpha_{\rm obs}$ is expected from the results of Sect. \ref{sec:1-spot} (see text for details).} \label{fig:2spot_LL}
\end{figure*}

The first source for false-positives/negatives for the sign of differential rotation was already identified from the 1-spot simulations. The method summarized in Sect. \ref{sec:method} is only fully valid for light curves whose spot modulation is induced by spots on the northern hemisphere. The method will also return the correct sign when the two spots are on opposite hemispheres, but only if the spot on the southern hemisphere is at a lower absolute latitude than the spot on the northern hemisphere. This is shown in Fig. \ref{fig:2spot_LL}, confirming that if the two surface rotation periods are successfully detected and distinguishable, the conclusions for 1-spot simulations will be valid for 2-spot simulations. Here, the sign of the surface differential rotation is determined by $\alpha_{\rm obs}$ (Eq. (\ref{eq:alpha})).

\begin{figure}[h]
\includegraphics[trim=11 0 6 0 mm,clip,width=\hsize]{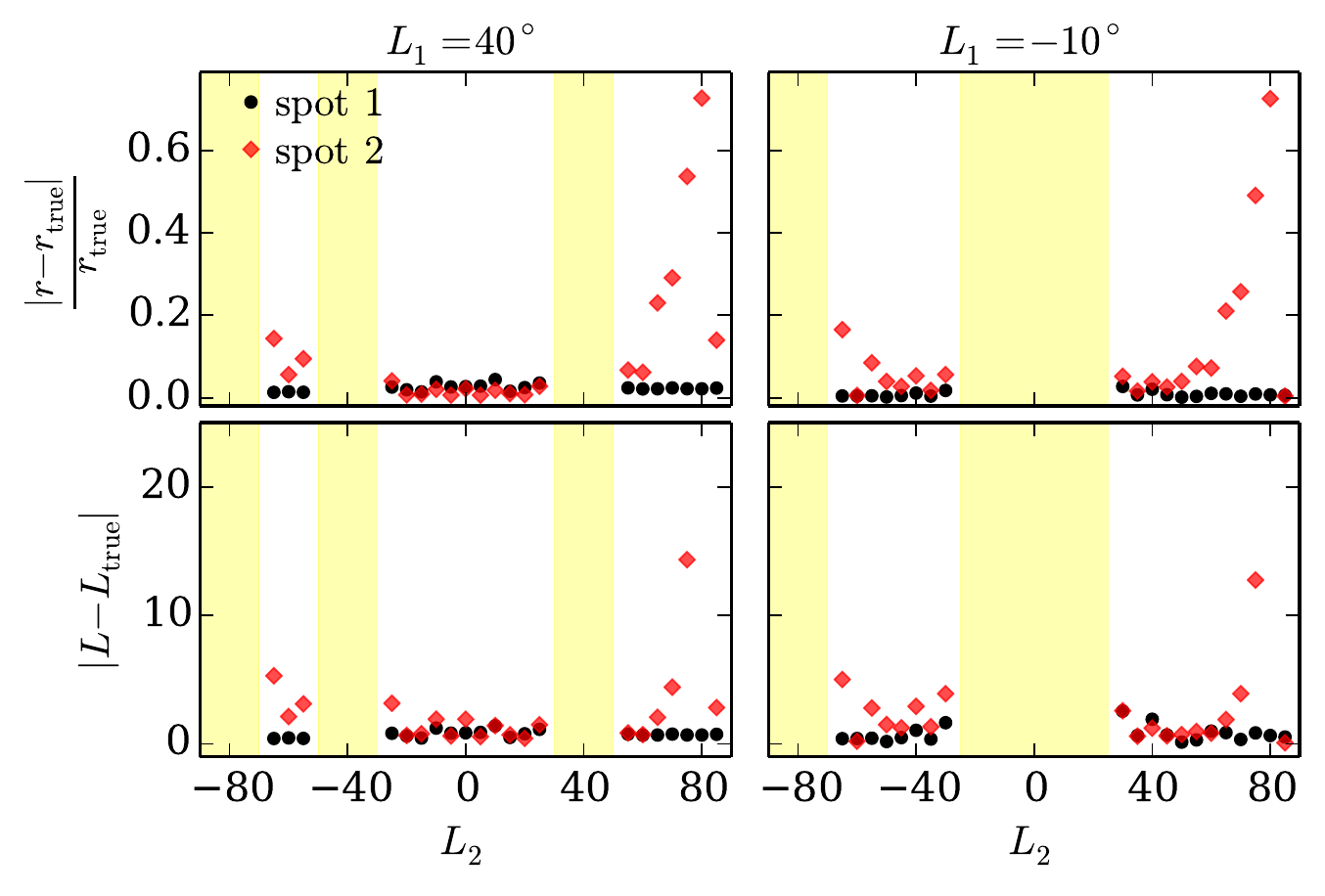}
\caption{Error on the peak-height ratios (top) and latitudes (bottom) as a function of the latitude of the second spot. Black and red symbols concern spots 1 and 2, respectively. The left panels correspond to $L_1\!\!=\!40^\circ$, while the right panels show the results for $L_1\!\!=\!-10^\circ$. The yellow areas mark the latitude intervals where only one rotation period was detected. Here, we disregarded the cases in which the peak-height ratios were outside the 1-spot peak-height ratio range given in Fig. \ref{fig:ratioL}.}\label{fig:2s_rL}
\end{figure}

Figure \ref{fig:2s_rL} shows the errors on the recovered peak-height ratios and inferred latitudes as a function of $L_2$ for two particular cases with $i=70^\circ$. Left and right panels correspond to $L_1=40^\circ$ and $L_1=-10^\circ$, respectively. The errors on the ratios are determined in relation to the reference values shown in Fig. \ref{fig:ratioL}. Taking the reference latitude-ratio relation and the peak-height ratios recovered from the 2-spot simulations, the "observed" spot latitudes $L$ can be inferred and then compared with the input latitudes. The yellow areas mark the latitude intervals where only one rotation period is successfully detected. For the cases shown, the error on the spot latitude is at maximum $\sim 15^\circ$. This indicates that, if the stellar inclination angle is known, the observed peak-height ratios, together with the results from 1-spot simulations (the latitude-ratio relation for the corresponding $i$), can be used to estimate the latitudinal distribution of spots.

\subsubsection{Spots' area effect}

\begin{figure}[h]
\includegraphics[width=\hsize]{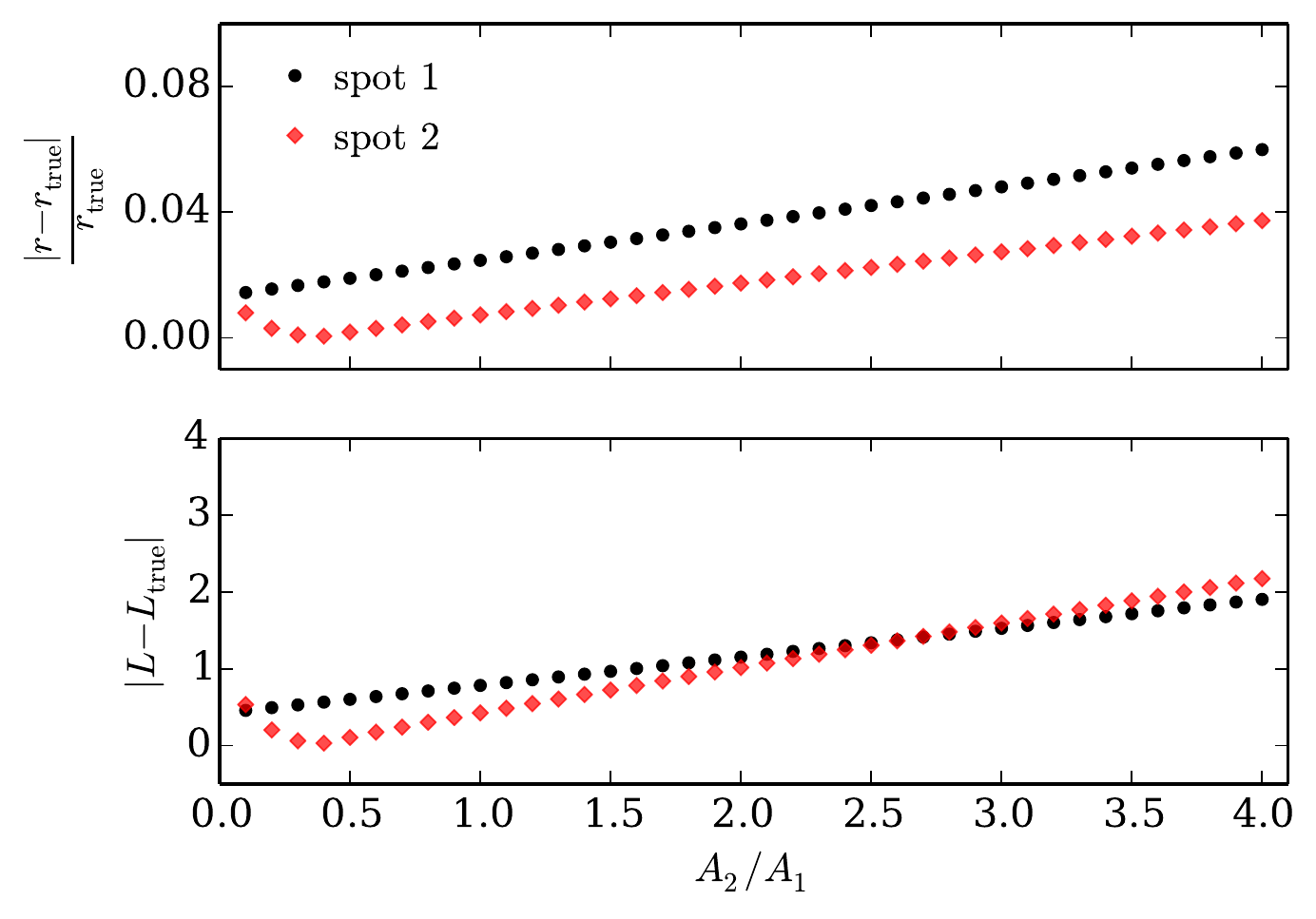}
\caption{Error on the peak-height ratios (top) and latitudes (bottom) as a function of the ratio between the areas of spot 2 (red) and spot 1 (black). For all the cases, $\alpha_{\rm obs}>0$.}\label{fig:2s_rA}
\end{figure}

Figure \ref{fig:2s_rA} shows the errors on the peak-height ratios and latitudes as a function of the spot area ratio, $A_2/A_1$. For this set of simulations, the spot latitudes ($L_1\!=\!40^\circ$ and $L_2\!=\!20^\circ$), stellar inclination angle ($i\!=\!70^\circ$), and the surface rotation ($\Omega_{\rm eq}\!=\!\Omega_{\rm eq,\odot}$, $\alpha\!=\!\alpha_{\odot}$) are fixed. The spots at $L_1\!=\!40^\circ$ have a constant radius of $R_{1}\!=\!5.7^\circ$, while the radius of the spots at $L_2\!=\!20^\circ$ varies between $1.8^\circ$ and $11.5^\circ$. The results show that the errors in the inferred peak-height ratios and latitudes are not significantly affected by variations in the relative area of the spots. In this case, a solar differential rotation ($\alpha_{\rm obs}>0$) is correctly recovered for all the synthetic light curves.

\subsubsection{Spots' phase effect}

\begin{figure}[h]
\includegraphics[width=\hsize]{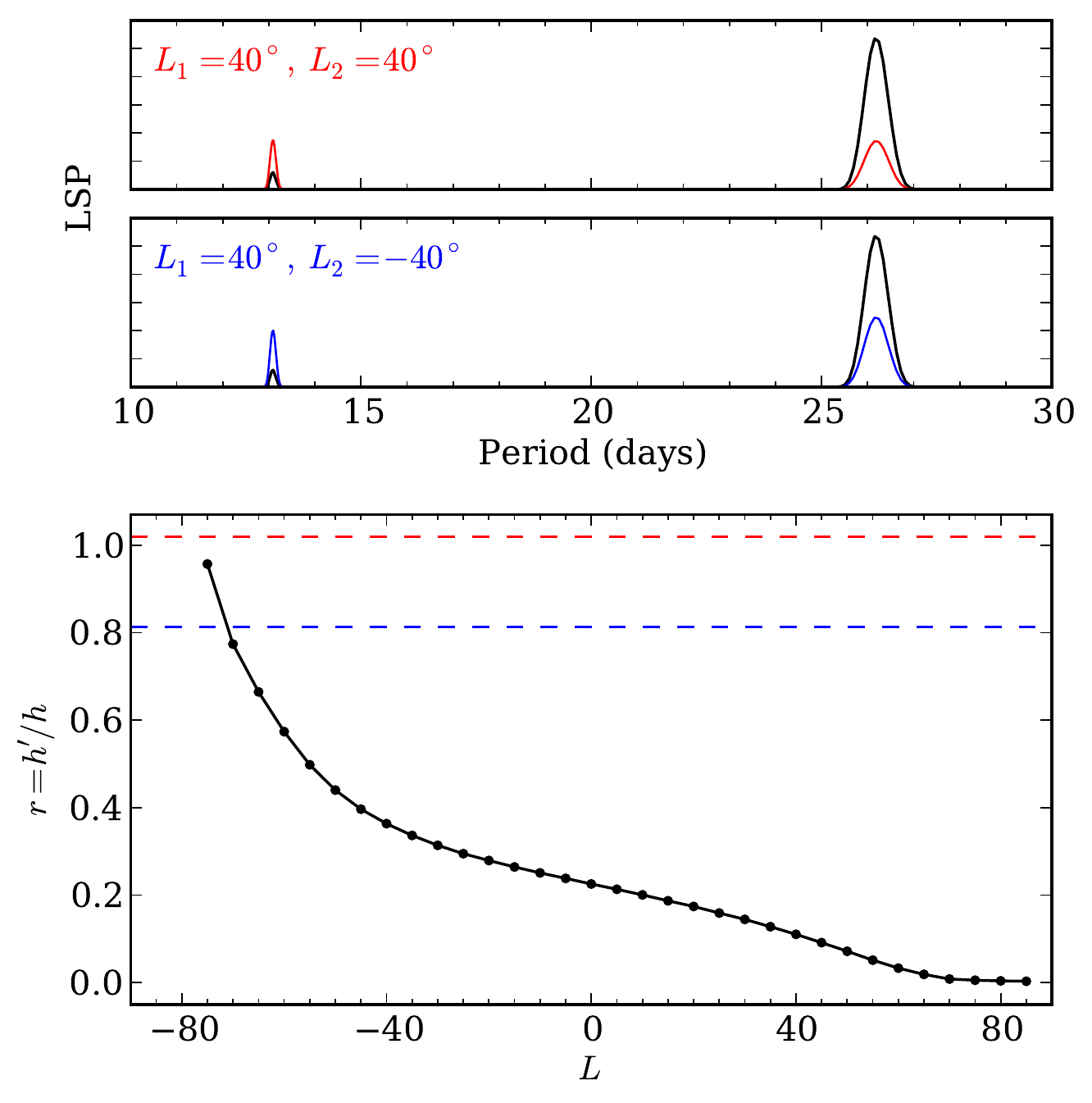}
\caption{Top and middle panels: Lomb-Scargle periodogram of two synthetic light curves whose modulation is produced by two spots rotating with equal velocity and in anti-phase. In the first example (red), both spots are on the northern hemisphere, while in the second panel (blue), the spots are on opposite hemispheres. The black line corresponds to the reference case of one spot at $L_1=40^\circ$. Bottom: Comparison between the peak-height ratios recovered from the top and middle panels (red and blue, respectively) and the reference ratios for $i=70^\circ$ (black).}\label{fig:LSP_phase}
\end{figure}

The phase of the spots also has an impact on the peak-height ratios. In particular, when spots have similar rotation rates and are in anti-phase, the LSP shows an excess of power on the second harmonic. In some cases the second harmonic can even be the main peak in the LSP, being wrongly identified as the rotation period of the star \citep[e.g.][]{McQuillan2013,Reinhold2013}. In these cases, the resulting peak-height ratios should not be used to infer the spot latitude or the sign of differential rotation. Two examples are shown in Fig. \ref{fig:LSP_phase}, where we consider a stellar inclination angle of $70^\circ$ and the spot latitudes $L_1=40^\circ$ and $L_2=\pm40^\circ$. When $L_2=-40^\circ$, the two spots have the same size (corresponding to $R_{\rm S}\sim5.7^\circ$), while when $L_2=40^\circ$, the second spot is half the size of the first spot\footnote{Note that if the spots have the same size and latitude, the modulations produced by each spot will have equal amplitude. In this case, the signature of the two spots rotating in anti-phase would be equivalent to the modulation of one spot rotating twice faster than the rotation period. This means that one would retrieve half of the rotation period.}. For comparison, the black line corresponds to the reference LSP for one spot at $40^\circ$, with a radius of $5.7^\circ$. In both cases (top panel - red and middle panel - blue), the observer could be wrongly led to assume that the peak in the LSP is being produced by a single spot, but the peak-height ratios in both cases would be very different from the case of a single spot (in black). This is also evident from the bottom panel which compares the recovered ratios with the reference ratios from Fig. \ref{fig:ratioL} for the inclination of $70^\circ$. If we still considered the higher period as the first harmonic, the peak-height-ratio that would be inferred in the first case (red) would be outside the expected range for a single spot for the chosen inclination, while in the second case (blue) one would infer a very low latitude, if the single spot scenario were to be wrongly assumed. The longitude of the second spot (in both examples) is $\phi_2=\phi_1+\pi$.

\begin{figure}
\includegraphics[width=\hsize]{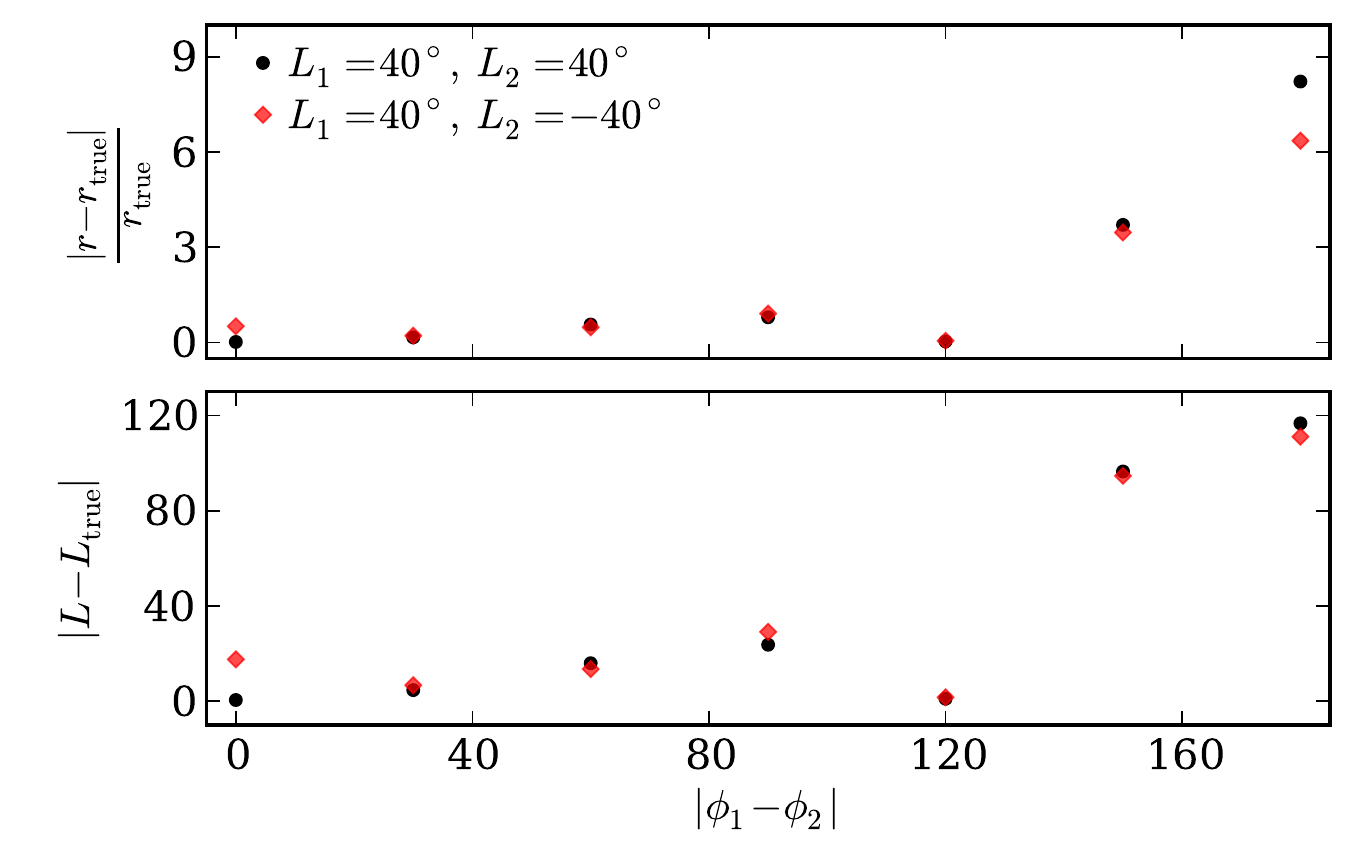}
\caption{Error on the peak-height ratios (top) and latitudes (bottom) as a function of the phase difference between two spots rotating with equal velocity, both on the northern hemisphere (black) and on opposite hemispheres (red). The reference case corresponds to the one in Fig. \ref{fig:LSP_phase} for a spot at $L=40^\circ$.}\label{fig:2s_rL_phase}
\end{figure}

Figure \ref{fig:2s_rL_phase} shows the error on the estimated peak-height ratios and inferred latitudes as a function of the phase difference between the two spots rotating with equal velocities (for the same latitudes of Fig. \ref{fig:LSP_phase}, $L_1=40^\circ$ and $L_2=\pm40^\circ$). Clearly, for certain phase differences the inferred latitudes and peak-height ratios would be far from the true values. The results in both Figs. \ref{fig:LSP_phase} and \ref{fig:2s_rL_phase} thus confirm that one needs to be cautious on the analysis of light curves showing evidence of spots rotating in anti-phase.

\vspace{0.5cm}
\section{Conclusions}\label{sec:conclusions}

The main goal of this work was to understand under which conditions the spot modulation on the light curve and its signature on the periodogram can provide insights into the latitudinal distribution of starspots and consequently, into stellar surface differential rotation. In particular, we studied the dependence of the peak-height ratios, computed from the periodogram, on the spot and stellar parameters.

We found that the peak-height ratios depend essentially on the fraction of time the spot is visible. Spots that are visible for longer time compared to $P_{\rm rot}$ produce smaller ratios than spots that are visible for a shorter time. In turn, the spot visibility time depends more significantly on the stellar inclination angle and spot latitude. 

Our results from 1-spot and 2-spot simulations show that, the method proposed by \citet{Reinhold2015} provides the wrong sign of surface differential rotation when:
\begin{itemize}
\item $i\neq90^\circ$ and the peak-height ratios are associated to spots on the opposite hemisphere of the observer;
\item $i\neq90^\circ$, one of the spots (spot 1) is on the opposite hemisphere while the second spot (spot 2) is on the same hemisphere of the observer and $|L_1|>|L_2|$;
\item the peak-height ratios are related to spots rotating with similar velocities and nearly in anti-phase.
\end{itemize}

Moreover, for low inclinations, the peak-height ratios become saturated as a result of spots being always visible for a wide range of latitudes. Also, for $i=90^\circ$ the peak-height ratios are almost constant. In these cases, attributing a latitude to each rotation period and determining the sign of differential rotation will be difficult.

Despite the degeneracy between stellar inclination angle and spot latitude, we find that the peak-height ratios provide a simple and fast way to constrain those parameters. This is a clear advantage of this method in comparison with other time consuming methods \citep[e.g.][]{Mosser2009,Huber2010,Walkowicz2013,Lanza2014}, where the inclination, spot latitude, area and intensity contrast may be strongly degenerated. Moreover, if the inclination angle is known, the peak-height ratios can constrain the latitudinal distribution of spots.

The spot signature on the light curves depends on a number of stellar and spot properties, such as the stellar surface rotation, limb-darkening law, spot size, and intensity contrast. We have investigated how the peak-height ratios depend on those parameters. We found that the effect of the spot size and limb-darkening on the peak-height ratios is small but not negligible.

We have also shown that, when two rotation periods are successfully recovered, the conclusions taken from the 1-spot simulations are also valid for 2-spot simulations. Moreover, although the relative size of the spots (for 2-spot simulations) affects the ratios, the effect is in general not strong enough to lead to a wrong inference of the sign of differential rotation. 

We have not considered spot evolution, which is beyond the scope of this study. However, we note that the multiple peaks in the periodogram can also result from spot evolution \citep[e.g.][]{Lanza2014,Aigrain2015,Reinhold2015a,Nagel2016}. For stars showing evidence of long-lived spot/active regions that induce stable signals, the LSP and the peak-height ratios will be less affected by the spot evolution. The analysis of different subseries of the full light curve may also help on discriminating between periodic (or quasi-periodic) signals related to the stellar rotation and those resulting from other sources.

Finally, we note that there is an observational bias, which contributes to the small number of false-positives reported in \citet{Reinhold2015}, since the modulation induced by spots on the same hemisphere of the observer will be preferentially observed in comparison with spots on the opposite hemisphere, in particular for small inclination angles. 

\begin{acknowledgements}
This work was supported by Funda\c{c}\~{a}o para a Ci\^{e}ncia e a Tecnologia (FCT) through the research grant UID/FIS/04434/2013. ARGS acknowledges the support from FCT through the Fellowship SFRH/BD/88032/2012 and from the University of Birmingham. MSC and PPA acknowledge support from FCT through the Investigador FCT Contracts No. IF/00894/2012 and IF/00863/2012 and POPH/FSE (EC) by FEDER funding through the programme Programa Operacional de Factores de Competitividade (COMPETE). RAG acknowledges the support of the GOLF and PLATO grants. SM would like to acknowledge support from NASA grants NNX12AE17G and NNX15AF13G and NSF grant AST-1411685. The research leading to these results has received funding from EC, under FP7, through the grant agreement FP7-SPACE-2012-312844 and PIRSES-GA-2010-269194. ARGS, MSC, and PPA are grateful for the support from the High Altitude Observatory (NCAR/UCAR), where part of the current work was developed.
\end{acknowledgements}

\bibliographystyle{aa}
\bibliography{diffrot}

\end{document}